# Semantic model for the description of energy data in the Module Type Package

Leif-Thore Reiche
*Institute of Automation Technology*
Helmut-Schmidt-University Hamburg
Hamburg, Germany
leif-thore.reiche@hsu-hh.de

Felix Gehlhoff
*Institute of Automation Technology*
Helmut-Schmidt-University Hamburg
Hamburg, Germany
felix.gehlhoff@hsu-hh.de

Alexander Fay
*Chair of Automation*
Ruhr University Bochum
Bochum, Germany
alexander.fay@rub.de

*Abstract* — Modular production systems that employ the Module Type Package (MTP) to describe module interfaces can, at present, only communicate energy data through proprietary solutions. Due to this limitation, users face additional effort when calculating energy KPIs for modules or determining the energy efficiency of modules. To address this issue, we present a model that facilitates energy data to be described semantically and uniformly in the MTP on the basis of an industrial standard (OPC 34100). MTPs incorporating this model can transmit semantically consistent energy data from modules to the process control system, making the data available for further applications, such as monitoring or optimization.

## I. Introduction

The ongoing climatic changes, the tense situation in the global economy and supply chains and the lack of skilled professionals have led to an increase in investment risks for the construction of automated industrial plants [1]. At the same time, system operators are faced with the problem of manufacturing new and increasingly customized products for volatile markets [2]. As a solution to these problems, the modular production approach has become widely accepted in recent years, which, according to the ZVEI, is expected to affect a quarter of newly planned industrial plants in the process industry by 2030 [3]. The modular production differs from the conventional approach in that a production is not planned and developed as a single monolith, but is made up of individual modules known as process equipment assemblies (PEA) [4]. In such a production plant, each PEA fulfills specific tasks and functions in the form of services, which together constitute a production process [5]. In order to control the PEAs with the help of a process control system, they are connected via a standardized interface, the Module Type Package (MTP), on a higher level, also known as the Process Orchestration Layer (POL) [6].

The increasing electrification of industry and fluctuations on the energy markets have increased the importance of energy management according to the ISO 50001 standard as well as the importance of energy data in industrial applications (e.g., for improving energy efficiency) [7, 8]. As energy data is not yet explicitly considered in the current status of the MTP, a semantic model for the description of energy data in the MTP is accordingly not available [9].

Without energy information it is not possible to perform energy management. If operators of modular plants want to access the energy data from PEAs, they are forced to develop individual solutions for recording and communicating the energy data. Such solutions are either based on the current specifications of the MTP (and, therefore, only provide an inadequate description of the energy data because of too little semantic content), or completely new proprietary solutions are developed.

In order to solve this problem, this article presents a description model that enables the semantically uniform description of energy data based on an industrial standard in the MTP. With the help of this model, energy data from different sources of a PEA can be communicated to the POL in order to be available for further data processing. The article takes up previous work such as [9, 10] and extends it.

This article is structured as follows: Section 2 presents the current state of the art and discusses related work. Section 3 presents a methodology used to generate the semantic model. Section 4 presents the semantic model for describing energy data in the Module Type Package. In order to explain how the model works, it is evaluated in Section 5 using an application example.

## II. State of the Art

To get an overview of the current state of the art this section gives an insight into the topics (A.) energy data in industrial applications (B.) modular production with the Module Type Package and (C.) elements for the communication of data within the Module Type Package.

### A. Energy data in industrial applications

Energy data is relevant whenever energy management should be executed within industrial applications [8]. For example, energy data is used as the basis for determining energy efficiency or other energy KPIs [11]. Before such calculations and evaluations of energy data can be started, the energy data must be recorded [12]. In industrial applications, energy data is usually recorded at the field level and communicated to the higher levels of the automation system, where it is then further processed. According to authors of [7], the OPC UA communication technology and various Ethernet-based and non-Ethernet-based communication protocols (such as PROFINET, Sercos, Modus) can be used to communicate energy data from the field level to higher levels of the automation system (e. g. process control system). The authors of [7] explain further that the applicable protocols have different semantics for the description of energy data. If several of these protocols will be used in an industrial application, the engineering effort increases when merging the energy data, because the raw data has to be converted, descriptions have to be changed, or data types have to be adapted. To solve this issue, the authors of [13] suggest to use an energy information model that provides a semantically standardized description of energy data. This recommendation was taken up by the industry so that a Joint Working Group was formed under the leadership of the German Verband Deutscher Maschinen- und Anlagenbau e.V. (VDMA), which is developing a companion specification for the description of semantically standardized energy data [14]. This standard can be used to aggregate energy data from various sources in industrial applications and describe it in a semantically uniform way

*B. Modular production with the Module Type Package*

Modular production is an approach in which a production process is composed of so-called process equipment assemblies (PEAs) [4]. In a modular production, the PEAs fulfill individual functions of the process (e.g. mixing, heating, filling) by providing so-called services [5]. In automation technology terms, each PEA is provided with a so-called Module Type Package (MTP) in order to standardize the interface between the PEAs and a higher-level process control system [4]. The MTP is a standardized description file of the PEA interfaces, which defines the functions and services of the PEA as well as the data to be exchanged and the human machine interface. For the description of the MTP, the AutomationML format based on the IEC 62424 standard (CAEX) is used [15]. If a PEA has an MTP, the MTP can be imported into a higher-level control system on the so-called process orchestration layer (POL) where it is orchestrated [6]. During the operation of modular production, it is possible to access the individual functions and services of the PEAs via the control system in order to map the relevant process data.

*C. Communication of energy data within the Module Type Package*

In order to be able to exchange data bidirectionally between a PEA and the POL via the MTP, the OPC UA communication technology is used. Part 3 of the VDI 2658 series of VDI standards [16] describes various data objects for this data exchange. For example, to control actuators in a PEA, different *ActiveElements* can be used, which describe the data sets for valves or motors. Furthermore, the descriptions in Part 3 also explain how data can be displayed using *IndicatorElements*. With the help of the *IndicatorElements*, it is possible to display analog values, integer values, binary values or even character strings, for example. Based on the analysis of [9], it can be concluded that the semantically uniform description of energy data has not yet been mentioned in this part of the guideline or in the associated parts of the guideline series.

Interim conclusion: As the current state of the art does not provide any solutions that can be used to communicate energy data from a PEA to the POL in a semantically standardized way, users are faced with the problem of designing their own proprietary solutions that contradict interoperable modular production. In order to solve this problem, this article proposes the use of a model which can be used to describe the energy data in a semantically standardized way. This is based on the fact that such models (e.g., [17, 18]) are already used in industry to describe energy data. The method described in the next section examines whether and to what degree these existing models can be used. The method is also used to generate the model for the semantically standardized description of energy data in the MTP.

### III. APPROACH TO CREATE A SEMANTIC MODEL FOR ENERGY DATA IN THE MODULE TYPE PACKAGE

In order to be able to provide semantically standardized energy data in modular production, a method is required that considers existing standards for the description of energy data and concepts for expanding the MTP, compares these with the current MTP standard, and makes recommendations for action based on the results. The steps of this method are listed below:

1. <u>Evaluation of the usability of existing semantic models for the description of energy data:</u>
   The first step of the method is to evaluate whether existing standardized models for the description of energy data can be reused and whether they can be embedded in the MTP. The reuse of existing and standardized models makes sense, as many domain experts have already contributed to the standards. Otherwise, a completely new model for energy data would have to be created in the MTP in addition to the existing models.

2. <u>Investigation of the extent to which energy data can be communicated using the existing MTP functions:</u>
   The second step of the method is used to investigate whether the existing options in the MTP are sufficient and whether they can be used to describe energy data in a semantically uniform way. Within the scope of this investigation, the existing data objects are considered in particular.

3. <u>Creation of the data object EnRGView</u>
   Step 3 of the method suggests the introduction of a new data object that is based on existing standards, and which can be used for describing energy data within the MTP. The new data object extend the current functionalities of the MTP.

4. <u>Embedding data objects in existing solutions</u>
   In order to be able to locate the new data object in the MTP's overall context, the placement of the object will be explained in step 4 of the method. This mapping is based on an existing concept [9] that already suggests an extension of the MTP.

### IV. SEMANTIC MODEL FOR THE DESCRIPTION OF ENERGY DATA IN THE MTP BASED ON EXISTING STANDARDS

To generate the semantic model for energy data, the results of the steps described in Section III are presented.

*A. Evaluation of the usability of existing semantic models for the description of energy data*

In order to make energy data from a PEA available to the POL via the MTP in a semantically standardized form, it is advisable to examine the integration of existing companion specifications for energy data as an option. The use of companion specifications makes sense, as these have already been developed and tested by experts in various committees. For example, the integration of the companion specification of the PROFINET-based energy profile PROFIenergy is conceivable. However, as this energy profile can only map electrical energy data, this companion specification is only partially suitable. In order to be able to display both electrical and non-electrical energy data, it is recommended to integrate the newly developed companion specification PCM [14]. As the MTP is not yet mentioned in this specification and the VDI/VDE/NAMUR 2658 guidelines do not yet contain any general information about using companion specifications, a proprietary solution would have to be developed that would also apply to companion specifications with a different focus. The integration of companion specifications for energy data is therefore not pursued further at this point. Nevertheless, companion specifications provide a useful basis for describing energy data in a semantically standardized way, therefore, the contents of the companion specification (see [14]) is taken into account in the further steps.

### B. Investigation of the extent to which energy data can be communicated using the existing MTP functions

The next step of the method is to investigate which existing data objects would be suitable for the representation of energy data in the MTP. As mentioned in the state of the art (Section 2), data objects of different types can be mapped with *IndicatorElements*. Only the *IndicatorElement AnaView* is suitable for displaying energy data of a PEA with the existing functions of the MTP. The use of other elements for displaying binary values (*BinView*) or character strings (*StringView*) is not suitable. If the *DIntView* element is used to display energy data, the problem occurs that only integer energy values can be displayed and thus no precise energy measurements. The use of the *AnaView* indicator element enables the representation of a value measured in the PEA as well as the representation of the corresponding measured value unit. In addition, measured value-specific minimum and maximum scale values can be transferred. Figure 1 shows the properties of *AnaView*.

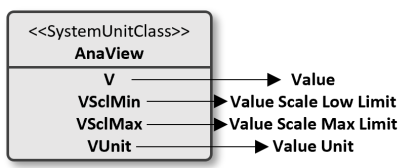

*Figure 1 AnaView*

The disadvantage of this solution is that important attributes of energy data are missing in this type of representation. For example, it is not possible to assign precise information about the performed measurement to the displayed analog value. A voltage measurement of phase L1 to neutral conductor N is only displayed as a value of 50V, but this measured value could also be a voltage measurement of phase L1 to phase L2. If such measurement information is stored in the measurement label (e.g. EI001VML1N - Energy meter 001 - Voltage Measurement L1-N), it can be assumed that the user is using proprietary semantics, which in turn contradicts the interoperable idea of standardized interfaces in modular production. It would also not be possible to communicate information on measurement accuracy for the measurement point when using *AnaView*. *AnaView* also offers no possibility to describe a measured resource. Whether a measurement is an electrical or non-electrical measurement and how it is characterized (e.g., non-electrical heat quantity) remains unanswered when using *AnaView*. In addition, no historical values from energy meters or measurement times can be stored in *AnaView*.

### C. Creation of the data object EnRGView

The findings from B. show that it is not yet possible to communicate energy data in a semantically uniform way with the given *IndicatorElements*. Based on this finding, an extension of the MTP-specific *IndicatorElement AnaView* is proposed in this step of the method. In addition, the contents of the companion specification mentioned in A. forms a basis for the extension, as energy data can be described semantically in a standardized way with the help of an energy information model in such a specification. Figure 2 shows the extended and inherited *AnaView* element with the name *EnRGView*. The figure also shows which contents of the energy information model of the companion specification can be adopted and which contents can already be represented by the contents of the *AnaView IndicatorElement*. The individual contents of *EnRGView* are briefly described below:

**MeasuringPoint[TagName /TagDescription]:**
To be able to describe a measuring point for measuring energy data, the existing *TagName* and the existing *TagDescription* can be taken from the Data Assembly class. The description of *TagName* and *TagDescription* can be compared with the application tag used in the energy information model of the companion specification.

**MeasurementValue [V]:**
In order to display an analog measurement value, the existing value in *AnaView* can be used. The value in *AnaView* is equivalent to the *MeasurementValue* from the energy information model of the companion specification.

**EngineeringUnit [VUnit]:**
The energy information model of the companion specification requires the usage of engineering units that are mapped via the struct data type. As measurement units are already contained in the *AnaView* data object, these are not adopted from the information model.

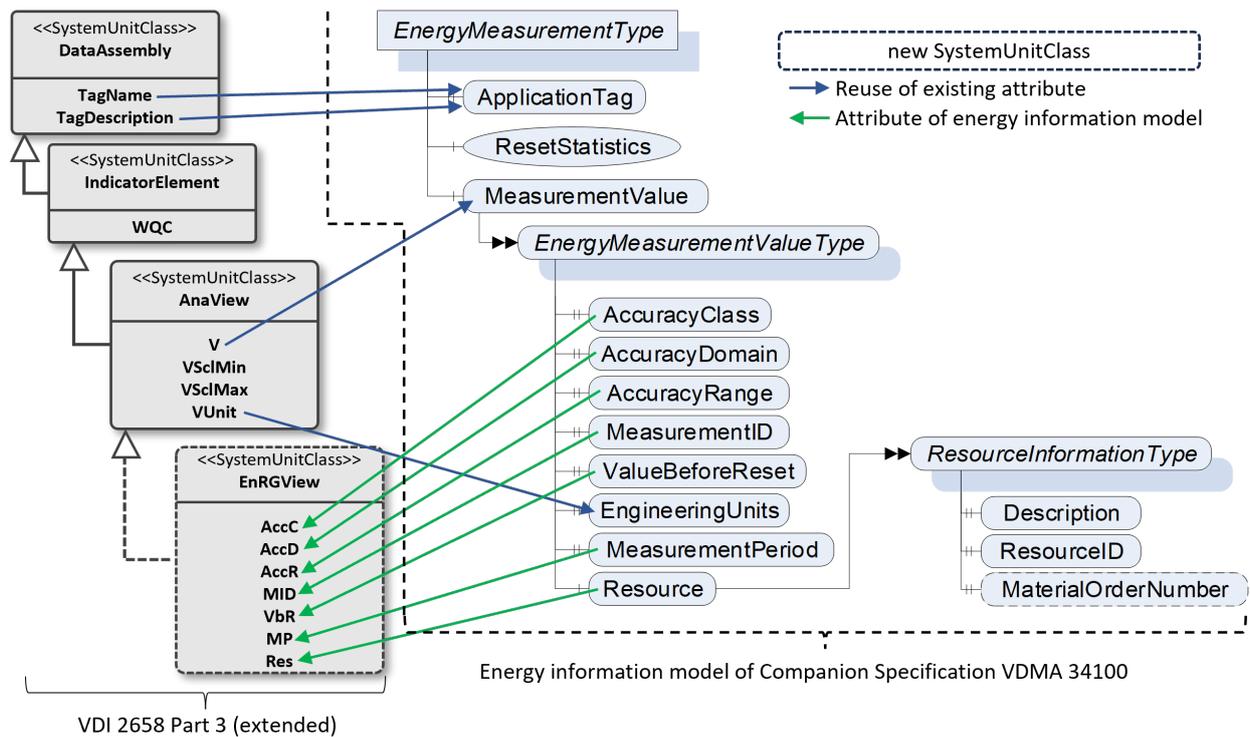

*Figure 2 Sematic model for energy data based on the energy information model of OPC 34100 [14]*

**Accuracy Class [AccC]:**
Specific classes are mentioned in the Companion specification for specifying measurement accuracies, with the help of which it is possible to make statements about the percentage deviations of the measurements. Since *AnaView* does not contain this attribute, it is included from the information model and inserted in *EnRGView*.

**AccuracyDomain [AccD]:**
Measurement accuracies can be specified further by using an accuracy domain. The domain is used to make statements about the extent to which the accuracy relates to the current measured value or to a series of measurements. *EnRGView* takes this attribute from the energy information model of the companion specification.

**AccuarcyRange [AccR]**
To display the full-scale value of a measured value, an accuracy range is specified in the energy information model of the companion specification.

**MeasurementID [MID]**
A measurement ID is specified in the energy information model to precisely identify an energy measurement. The ID can be used to decode the energy measurement via a table in the companion specification. This attribute is also adopted by *EnRGView*.

**ValueBeforeReset[VbR]:**
With *ValueBeforeReset* it is possible to keep values after resetting energy meters. This attribute is adopted by *EnRGView*.

**MeasurementPeriod[MP]:**
*MeasurementPeriod* can be used to specify periods for energy measurements. This attribute is also adopted by *EnRGView*.

**Resource [Res]:**
In order to be able to specify which energy resource is being measured (e.g. electricity, gas, coal), a resource is stored in the energy information model. The resource can also be relevant for measurements in PEAs and is therefore included in *EnRGView*.

### D. Embedding in existing solutions

The next step in the method is to localize the newly designed data object *EnRGView* in the MTP. The concept proposed in [9] will be used for this purpose. The concept extends the existing MTP with an additional aspect for energy management. With the help of the concept, it is possible to assign measurements to measuring points in the field. The new *IndicatorElement EnRGView* provides a semantically standardized description of the measurements of these measuring points. If the measurement of a measuring point should be described in the MTP, a so-called *MeasurementType* must be created for the measurement according to the concept. A *MeasurementType* corresponds to a measuring point instance and can contain several *EnRGViews* to ensure that, for example, several measurements can be assigned to a compact energy meter. The energy-specific measurement instances must be described for each *EnRGView* in accordance with IEC 62424 [19]. To further specify the measurements on the measuring device, the *MeasurementTypes* are grouped into specific module, service or component measurements and stored in a *MeasurementList*. In the MTP, this list is stored in the energy management functions. The functions in turn are subordinate to the energy management aspect. With the help of this structure, it is possible to uniformly locate the semantic information of an energy measurement.

## V. EVALUATION OF THE SEMANTIC MODEL FOR THE DESCRIPTION OF STANDARDISED ENERGY DATA

The next step is to evaluate the semantic model presented in Section IV. A PEA for distillation purposes (laboratory PEA) that has a compact energy meter is used as an evaluation example. The evaluation on this PEA is carried out in five steps, which are explained individually below:

1) Generating the MTP of the distillation PEA:
The first step in evaluating the semantic model is to generate an MTP for the distillation PEA using manufacturer-specific software. In the case of the distillation PEA, TwinCat3 from Beckhoff Automation has been used. The generated MTP is used as the basis for the subsequent work.

2) Extension of the MTP library to include the semantic model for energy data:
In the second step, the generated .mtp file is used and the contained manifest is opened with an AutomationML editor. Next, the semantic model is added to the *SystemUnitClass* library in the manifest. To do this, another *SystemUnitClass* called *EnRGView* is added to the *AnaView* data object under *MTPDataObjectSUCLib/DataAssembly/IndicatorElement*. The *SystemUnitClass EnRGView* is then extended by the attributes mentioned in ection IV and the data type and description are added for each attribute. In order to be able to place the energy data in the context of energy management, the library must be extended by an *MTPEnergyManagementSUCLib* according to [10]. Figure 3 shows the extended library in which the newly created data object and the associated attributes as well as the *MTPEnergyManagementSUCLib* can be seen.

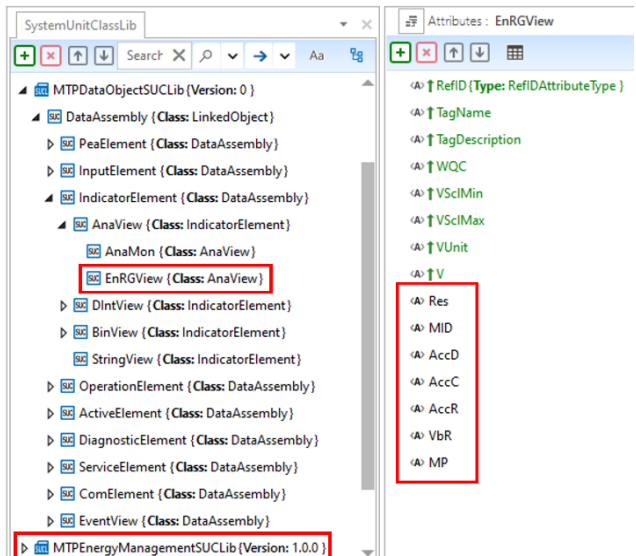

*Figure 3 Extension of MTP-Lib*

3) Extension of the MTP instance:
In the third step, the extended MTP library can be used to add energy data to the MTP instance. To do this, the individual measurement instances of the energy measurement devices can be created under *ModuleTypePackage/PEA/CommunicationSet/Instance*

*List* using the *EnRGView*. The attribute type of the *EnRGView* attributes is then assigned with an *IDLinkAttributeType*. Once this work has been completed, the OPCUA items of the individual attributes can be added to the OPCUA server in the *SourceList*. Accordingly, one *OPCUAItem* per *EnRGView* attribute must be added for each measurement instance. Once this work has been completed, the values and default values of the *OPCUAItems* can be mapped to instances in the OPC UA server. In addition, the link between the header of the *OPCUAItems* and the attributes of the *InstanceList* must be created via the *GUID*. After completing this work, the measurement instances are contextualized with the representation explained in [10].

4) Orchestration of the PEA on the POL
Once the semantic model for energy data has been added to the MTP, the project-specific POL engineering can begin in the next step. To do this, the MTP is loaded into a process control system. If custom faceplates should be used to display the energy data, these must be added to the HMI and linked to the *EnRGView DataAssembly*. The MTP can then be assigned to a device instance and the PEA can be orchestrated. When this step has been completed, the physical connection to the PEA must be set up and the communication with the PEA must be checked. If the communication is successful, the runtime of the process control system can be started.

5) Operation of the PEA:
When all activities have been completed, energy data can be accessed from each PEA of the modular plant. The semantically standardized energy data of the various measuring instances can then be viewed in the POL. Figure 4 shows the voltage value of the distillation PEA in the POL. This data can be easily opened using a separate faceplate. The energy data can also be provided as a table.

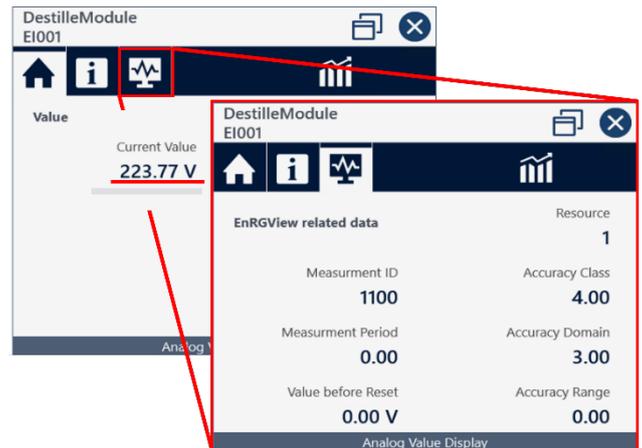

*Figure 4 Faceplate to show energy data (Voltage) in a POL*

The visualization of the energy data in the OL shows that the evaluation was successful, and the full functionality of the semantic model can be proven. As the work in this evaluation was carried out manually, it is recommended that the semantic model for energy data should be integrated into the manufacturer-specific MTP generators in the future. This would give users the advantage of not having to integrate the measuring points retrospectively.

## VI. Conclusion and future work

This paper has shown that the current scope of the MTP is not sufficient to transfer energy data from a PEA to the POL. One of the reasons is that the MTP does not yet have any descriptions that enables a semantically standardized description of energy data. This issue can also not be solved by using existing companion specifications for energy data, as the MTP interface cannot be used for the integration of a companion specification and therefore cannot transmit any energy data. This paper therefore presents a model based on industrial standards that enables energy data to be represented semantically uniformly in the MTP. The functionality of this model was demonstrated and explained using a laboratory example.

Future work will investigate the extent to which data of the POL can be used to control the energy demand of PEAs and how this can contribute to reducing the energy consumptions of PEAs.